# Are UX Evaluation Methods Truly Accessible to Deaf Users? An Empirical Study

# ¿Son realmente accesibles los métodos de evaluación de Experiencia de usuario para personas sordas? Un estudio empírico

Andrés Eduardo Fuentes-Cortázar[1], Alejandra Rivera-Hernández[2], José Rafael Rojano-Cáceres[3]

[1] Facultad de Estadística e Informática, Universidad Veracruzana.
Xalapa, Veracruz. México.
afuentescortazar@gmail.com
0000-0002-4620-6106

[2] Facultad de Estadística e Informática, Universidad Veracruzana.
Xalapa, Veracruz. México.
alejandra.rihe99@gmail.com
0009-0004-8739-8601

[3] Facultad de Estadística e Informática, Universidad Veracruzana.
Xalapa, Veracruz. México.
rrojano@uv.mx
0000-0002-3878-4571

Correspondence: rrojano@uv.mx[3]

**Abstract**

Providing an equitable and inclusive user experience (UX) for people with disabilities (PWD) is a central goal of accessible design. In the specific case of Deaf users, whose hearing impairments impact language development and communication, it is essential to consider their specific needs during software evaluation processes. This study aimed to analyze a set of UX evaluation methods suggested in the literature as suitable for Deaf individuals, with the goal of validating their level of accessibility in real-world contexts. The research was based on a critical review and practical application of these methods, identifying their strengths and limitations in relation to the interaction, perception, and comprehension of Deaf users. Traditional evaluation instruments, commonly designed for hearing individuals, pose significant barriers when applied to Deaf users due to their reliance on auditory and cognitive abilities, as well as the lack of consideration for communicational accessibility. The results show that although these methods are frequently recommended, they exhibit critical shortcomings that hinder the collection of accurate and representative data. It is concluded that it is essential to adapt UX evaluation methods to ensure genuinely accessible processes that address the communicative and cognitive needs of the Deaf community and accurately reflect their user experience.

**Resumen**

Proporcionar una experiencia de usuario (UX) equitativa e inclusiva para personas con discapacidad (PCD) es un objetivo central del diseño accesible. En el caso particular de los usuarios Sordos, cuyas limitaciones auditivas afectan el desarrollo del lenguaje y la comunicación, es fundamental considerar sus necesidades específicas durante los procesos de evaluación de software. Este estudio tuvo como objetivo analizar un conjunto de métodos de evaluación de UX sugeridos por la literatura como adecuados para personas Sordas, con el fin de validar su nivel de accesibilidad en contextos reales. La investigación se basó en la revisión crítica y aplicación de estos métodos, identificando sus fortalezas y limitaciones en función de la interacción, percepción y comprensión de los usuarios Sordos. Los métodos de evaluación tradicionales, comúnmente diseñados para personas oyentes, presentan barreras significativas cuando se aplican a usuarios Sordos, debido a su dependencia de capacidades auditivas, cognitivas y a la falta de consideraciones sobre accesibilidad comunicacional. Los resultados evidencian que, aunque estos métodos son frecuentemente recomendados, presentan deficiencias importantes que dificultan la recolección de datos precisos y representativos. Se concluye que es imprescindible modificar los métodos de evaluación de UX para garantizar procesos concretamente accesibles, que respondan a las necesidades comunicativas y cognitivas de la comunidad Sorda y reflejen de forma precisa su experiencia de usuario.

**Palabras clave**

Experiencia de usuario; Sordera; Accesibilidad; Evaluación; Desarrollo de software.

## 1 INTRODUCTION

Among the different types of disabilities, deafness is one that goes unnoticed in society [1] because it only becomes visible when individuals interact through their own language (sign language). However, while there are other invisible disabilities, such as mental disorders, deafness affects the way people relate and communicate with the rest of the non-deaf population, this is due to a cultural and linguistic contrast.

In particular, society has developed various verbal and written communication mechanisms that are often imposed on the deaf population without considering their specific communication and cultural needs.

As a result, it is worth asking whether and to what extent user experience (UX) evaluation methods reflect the expected measurements that are designed for people who use oral languages. Therefore, to evaluate deaf users with typical UX instruments, it is considered necessary to make proposals that adapt to their linguistic needs. As an example, it could be the adaptation of the SUS questionnaire to sign language [2]. Such an instrument is widely used to capture user usability in the Human-Computer Interaction (HCI) field.

Therefore, this paper explores the application of UX evaluation methods to users with hearing impairments who are sign language users. It is worth noting that this work is an extension of the study presented at the Ibero-American Conference on Human-Computer Interaction 2025 [3]. Firstly, the evaluation of UX methods is primarily reported as a case study, but now it addresses the details of the evaluation methodology in depth, and adds a contrasting study with other users who live with other disabilities to establish whether there are significant differences between the two groups. Likewise, the results are processed through statistical analysis in order to facilitate meaningful comparisons between different aspects observed during the evaluation. The evaluation rubric used to assess each participant's level of proficiency in Sign Language is presented below. Furthermore, the methodology and planning are developed with the intention of being evaluated, critiqued, and replicated by those interested in usability evaluation for deaf people.

Thus, the article is organized as follows: The current section presents the introduction. Section 2, describes the methodological approach, the population, the phases for conducting the research and the materials. Section 3, presents the UX methods selected to apply con deaf users. Section 4, describes the planning activities, which are the core of the study. Section 5, presents the results from applying the UX methods. Sections 6, presents the discussion, where each finding about the methods is detailed. Section 7, presents the conclusion and future work. Finally, we present the reference in section 8.

## 2 MATERIALS AND METHODS

The methodological approach adopted in this research is empirical in nature, as it prioritizes direct observation, active participation, and interaction with users as fundamental means for knowledge construction [4]. This methodological choice becomes particularly relevant when working with populations that have specific accessibility needs, such as Deaf individuals. The experiences of these users, along with their modes of communication and interaction with technological systems, differ significantly from the normative assumptions underlying traditional usability frameworks. Therefore, it is essential to adopt an approach that can accurately capture their perceptions, needs, and unique ways of interacting.

This study employs a quantitative/qualitative approach with an exploratory, cross-sectional design, where participants were selected through convenience sampling, given the specific physical characteristics required for the study's objective. Specifically, deaf students from a special education center in Mexico having these characteristics were recruited, the limited access to a large sample, justified the use of this type of non-probability sampling.

### 2.1 POPULATION

The study's target population consisted of people with prelingual hearing impairments who use sign language. As previously said, convenience sampling was chosen.

For this purpose, a sample of three young people was recruited from the Multiple Attention Center (CAM by its acronym in Spanish), located in a city within Mexican territory. This educational institution specializes in serving children and youth with disabilities, which may pose significant barriers to their inclusion in regular schools within the conventional education system.

The CAM provides an adapted environment and specialized resources that address the specific needs of this student population. The pedagogical work carried out by the professionals at this center aligns with the curriculum and academic programs currently in effect in Mexico [5]. It is important to note that students with various types of disabilities may coexist within the same classroom group, which requires differentiated educational planning and individualized support strategies to promote their comprehensive development.

Finally, it is worth mentioning that for this type of center, the consent of parents or guardians is already provided to allow interaction, taking pictures or conduct research with children.

## 2.2 DEMOGRAPHY

Three deaf children—two girls and one boy—participated in the present study, as previously mentioned with the authorization of their legal guardians. The participants ranged in age from 7 to 16 years and all had profound prelingual deafness, meaning it developed before language acquisition.

According to the Mexican educational system, two of the deaf participants were enrolled in primary education, while only one participant was attending secondary school.

During the evaluation period, other students who did not have a hearing impairment and were enrolled in primary education, shared the same classroom. These students exhibited other types of disabilities.

For the purposes of this study, participants were classified into two groups: group A, consisting of Deaf students, and group B, composed of students with other disabilities. **Table 1** provides a detailed breakdown of each participant's demographic information, including type of disability, age, and group belonging.

Table 1. Demography by user group. Source: Authors.

| Group | Disability | Gender | Age |
|---|---|---|---|
| (A) Deaf users | Deafness | Girl | 16 |
|  | Deafness | Girl | 7 |
|  | Deafness | Boy | 13 |
| (B) Users with other disabilities | Motor disability | Girl | 10 |
|  | Psychosocial disability | Boy | 9 |

As part of the inclusion, Mexican Sign Language (MSL) instruction is provided with a comprehensive approach encompassing grammatical, syntactic, and vocabulary aspects for all students enrolled. Within the classroom, a wide range of visual teaching materials are specially designed to facilitate learning for deaf students. These materials include graphic representations of the alphabet, days of the week, months of the year, names of school supplies, and animals, among other everyday concepts. Each element is presented clearly and visually, allowing for a direct connection between the sign and its meaning, thereby contributing to the progressive development of communicative competence in sign language for all participants.

## 2.3 SIGN LANGUAGE PROFILE

It is essential to identify the particular characteristics of each Deaf participant, as they present a range of conditions involving different intellectual and academic abilities and limitations. This heterogeneity requires differentiated attention that takes into account their spe-cific learning and communication needs. To establish an accurate diagnostic baseline, a preliminary assessment of the students' linguistic competence in SL was conducted by the teachers responsible for each group. For such purposes, two rubrics are applied to deaf students. In **Table 2** and **Table 3** are shown how teachers assign a SL level considering different aspects on the expression and communication of the deaf, as well as elements related to social interactions.

Table 2. Understanding, attitude and participation rubric. Source: Authors.

| ADVANCED Handles it with ease | INTERMEDIATE Requires consolidation | ENOUGH Requires practice | INSUFFICIENT Doesn't handle it |
|---|---|---|---|
| The student demonstrates confidence in using MSL. His or her body language and facial expressions are complete and demonstrate what he or she wants to communicate: happiness, sadness, annoyance, attention, calmness, affection, etc. | The student shows some insecurity in the use of MSL or, conversely, may become exaggerated, and may not be as clear in his or her expressions to communicate: happiness, sadness, annoyance, attention, calmness, affection, etc. | The student, with his expressions, shows nervousness, tension or seriousness, in the use of MSL, which changes the intention when communicating: joy, sadness, annoyance, attention, tranquility, affection, etc. | The student shows insecurity, sometimes even raising his voice or exaggerating the gestures he makes with his mouth, which makes it difficult to fully understand his expressions. |
| The student shows interest in the sender-receiver relationship. He or she stays focused on the topic under discussion. | The student shows interest in the sender-receiver relationship. He or she often strays from the topic because he or she doesn't fully understand what is being said. | The student displays a difficult attitude in the sender-receiver relationship. Despite trying, he or she doesn't always manage to stay on topic. | The student shows a lack of interest in the sender-receiver relationship. He or she is unable to participate in the sender-receiver relationship. |
| The student responds appropriately. Initiates a turn in the conversation. He or she uses appropriate gestures and signs to indicate that he or she is giving or taking a turn. | The student almost always responds promptly. He or she initiates turns in conversation. Occasionally, he or she fails to use correct gestures and signs indicating that he or she is giving or taking a turn. | The student has difficulty responding appropriately and initiating a conversation. He or she forgets to use correct gestures and signs to indicate whether he or she is giving or taking a turn. | The student gets lost in taking turns in conversation. He or she waits for someone else to signal that it's their turn, and often even avoids it. |

Table 3. Fluency expression, conversation and sign word rubric. Source: Authors.

| ADVANCED Handles it with ease | INTERMEDIATE Requires consolidation | ENOUGH Requires practice | INSUFFICIENT Doesn't handle it |
|---|---|---|---|
| The student uses MSL continuously and without pause in a conversation or presentation of a topic. He or she also uses fingerspelling for proper nouns and unknown words or words without ideograms in the same way. | The student uses interrupted conversation, which prompts comments from the recipient, but still manages to finish the discussion. His or her fingerspelling is either very slow or very fast, or he or she may even omit letters. | The student uses severely interrupted conversation, which leads to constant commentary from the recipient, an act that risks diverting interest and losing communication. Fingerspelling is inaccurate and can lead to confusing letters. | The student does not conduct a conversation and fingerspelling does not occur, as the student is slow, interrupted, hesitates, forgets ideograms, and makes unnecessary pauses. |
| The student has a precise command of signs in terms of configuration, orientation, movement, placement and expression. | The student occasionally uses signs with precision in terms of configuration, orientation, movement, placement, and expression. | The student lacks precision with signs of configuration, orientation, movement, placement and expression. Oc- | The student has difficulty signing in terms of configuration, orientation, movement, placement and expression. He |

| | | casionally, he or she confuses them with other ideograms. For example, he or she uses the same sign for different meanings. | or she confuses ideograms. |
|---|---|---|---|
| The student responds appropriately. Initiates a turn in the conversation. | The student almost always responds promptly. He or she regularly initiates conversational turns. | The student has difficulty responding appropriately and initiating conversations. He or she needs frequent support. | The student gets lost in taking turns in conversation. He or she even avoids it. |

The aforementioned rubrics were designed to assess the level of proficiency in Mexican Sign Language, in order to provide a more accurate and detailed view of each user's skills. This information is key to guiding and personalizing the teaching-learning process in the classroom, ensuring it aligns with the specific needs of each student. As a result **Table 4** provides relevant data about the deaf participants sign language skills.

**Table 4.** Fluency expression, conversation and sign word rubric. Source: Authors.

| User | Gender | Grade level | Sign Language level |
|---|---|---|---|
| 1 | Girl | 9th grade | Advanced |
| 2 | Girl | 2nd grade | Intermediate |
| 3 | Boy | 4rd grade | Insufficient |

### 2.4 Materials

To carry out data collection, several sessions were held with participants directly interacting with a mobile app called Signa App. This mobile application aims to facilitate the creation and edition using Signwriting notation [6], which is a pictographic notation to facilitate representation of sign language.

It was not intended to evaluate the app itself, but rather to leverage its features as a representative case to explore evaluation methods sensitive to the cognitive, linguistic, and cultural specificities of deaf people. Its use in this study generated evidence on the relevance of adapting both the instruments and the UX evaluation dynamics to the visual-gestural communicative context of this community. A representative screen from the app can be seen in **Figure 1.**

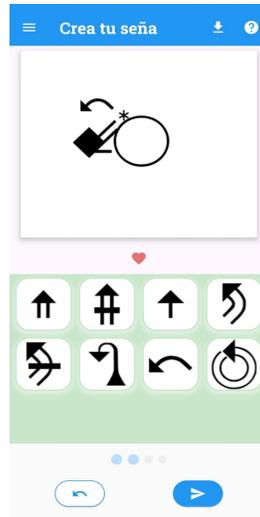

**Figure 1.** Screen "Create your sign". Show the sign "hello" in MLS. Source: Authors.

## 2.5 Methodology

Aiming to ensure a systematic and rigorous development of the research, and in order to allow replication of the study, a set of phases were defined. These respond to both methodological and practical criteria, considering the specific objectives of the study, the particular characteristics of the participant population, as well as the authors' prior experience in related research, were all taken into consideration. Based on these considerations, the following stages of the research process were defined:

1. **Population Selection:** In this initial phase, the target group—composed of Deaf users—was identified and delineated. The choice of this population was justified based on the central objective of the research, which aims to evaluate user experience from an accessibility-centered perspective. To ensure methodological relevance, the communicative and cognitive particularities inherent to this group were taken into consideration, recognizing the need to adapt procedures to their specific contexts. This consideration was fundamental to guaranteeing that the employed strategies effectively addressed the actual conditions and needs of the participants, thereby promoting an inclusive and representative evaluation. Taking this in consideration, the mobile application Signa App [7] was chosen because it was developed considering the deaf users characteristics.

2. **Formation of the Evaluation Team:** During this stage, an evaluation team was assembled, consisting of three researchers and one mediator. The researchers, all with experience in assessments involving deaf users, served as evaluators, contributing their technical and methodological expertise to the process. The role of mediator was undertaken by a teacher who, in addition to being part of the pedagogical team, acted as an expert in the educational field. Her participation was essential, as she fulfilled a dual role: on the one hand, she facilitated communication between the technical team and the participants; on the other, she provided accurate interpretation from both cultural and linguistic perspectives. This mediation ensured that all interactions were understandable, culturally appropriate, and aligned with the specific needs of deaf users throughout the entire evaluation process.

3. **Selection of UX Evaluation Methods:** In this phase, UX evaluation methods specifically documented in the literature for use with deaf users were selected. This selection was based on a prior literature review, as well as on recommendations drawn from similar studies conducted in comparable contexts. Additionally, criteria related to the specific characteristics of the participants were taken into account, including age, cognitive abilities, linguistic skills, and the level of engagement they could assume during the evaluation. The suitability of each method was carefully assessed in light of these factors. In this process, the input of the teacher-mediator was crucial, as her knowledge of the group enabled the contextualization and adaptation of the selected methods to the specific needs of the participating children, thereby ensuring a more relevant and effective application within the evaluation environment.
4. **Planning of Tests and Activities:** Based on the previously selected evaluation methods, a detailed plan was developed outlining the tests and activities to be conducted throughout the research process. This plan included an organized sequence of tasks, the required materials, estimated execution times, and the necessary support strategies to facilitate participant engagement. The fundamental purpose of this planning was to design an accessible and motivating environment that encouraged active interaction from the participants and enabled the collection of meaningful data regarding their user experience with the application. In this way, it was ensured that the activities were not only appropriate for the characteristics of the group but also conducive to obtaining relevant and reliable results.
5. **Execution of Assessment and Data Collection:** In this final phase, the evaluations were conducted in a controlled environment under the continuous supervision of the research team and the teacher-mediator. During this stage, meticulous observation of the users' interactions with the mobile application was carried out, systematically recording their behaviors, reactions, and comments. Concurrently, qualitative data were collected to support subsequent analysis, thereby enabling a comprehensive and detailed understanding of the participants' experience with the evaluated system.

After defining the distinct phases that structure this research, the following sections will focus on a detailed analysis of the implementation of various user experience evaluation methods specifically applied to Deaf users.

## 3 UX METHODS FOCUSED ON DEAF USERS

In this research, five UX evaluation methods were selected with the aim of applying them to Deaf individuals, on the premise that they may adapt to the unique needs of this population, ensuring both effective and inclusive evaluation processes.

The main objectives were: a) to obtain a comprehensive understanding of their user experience, and b) to analyze the level of accessibility provided by each of the methods used in the study.

The selection of each evaluation method was based on a review of specialized literature [8], which helped us to identify instruments previously used in software testing with Deaf users. This process aimed to ensure the relevance and suitability of the proposed methods within the study's context. Additionally, to guarantee the appropriateness of the selected techniques, these methods were later presented to the teacher responsible for the participants. This allowed the incorporation of the teacher's knowledge regarding the participants' cognitive abilities, as well as their recommendations on methods that could encourage active and meaningful participation.

### 3.1 User workshop

This method aligns with a participatory design approach, in which potential users collaborate with the design team during structured sessions. These workshops can be conducted either in person or virtually, offering flexibility in terms of participation and logistics. Although organizing such sessions requires careful planning and resource investment, they are particularly effective for gathering meaningful feedback and gaining valuable insights directly from users [9].

During these sessions, participants interact with the design concept, share their opinions on the proposal, and suggest potential improvements to the product or service being evaluated. The activities are led by facilitators from the design team, who are responsible for guiding the process, encouraging active participation, and ensuring that the workshop objectives are effectively met.

### 3.2 Direct observation

Direct observation is a purely observational research technique in which the researcher limits themselves to watching the participants' behavior without intervening or interacting with them. This methodology is especially useful for studying interaction processes and usability, as it allows the recording of behaviors in natural contexts, thereby providing a more authentic and accurate view of users' real practices, habits, and needs [10].

Since participants are not influenced by the active presence of the researcher, they tend to act spontaneously, which helps gather more objective and less biased data. This type of observation is particularly valuable in environments where it is necessary to understand how users interact with a product or system in everyday situations.

### 3.3 Cognitive walkthrough

The cognitive walkthrough is a usability-focused evaluation technique that analyzes how easy it is for users to learn to use a system or interface, particularly during their initial interactions with it. This method simulates the experience of a novice user, with the aim of identifying elements of the interface that might cause confusion, errors, or difficulties during the learning process [11].

During its application, the researcher assigns specific tasks to participants, who are asked to carry them out by following a logical sequence of actions. Throughout the activity, both the time taken to complete each task and any errors made are recorded. This allows for an assessment of the system's effectiveness and efficiency from a cognitive perspective. This technique is especially useful, as it provides valuable insights into potential comprehension barriers and opportunities to improve the user's initial experience.

### 3.4 Picture card through prototyping

This is a participatory technique that uses cards or cutouts containing images, keywords, graphic elements, and figures related to the product or service being evaluated. Its main purpose is to encourage narrative expression from participants, who use the cards as prompts to tell stories based on their past, current, or expected experiences with the product. The technique can also be applied to help users envision and represent their ideal version of the final product [9].

The cards are designed to depict both present situations and future scenarios, allowing for the exploration of user perceptions, expectations, and potential improvements. This technique is highly flexible, as both the content and the structure of the activity can be adapted to suit the characteristics of the participant group or the specific goals of the study. As part of the process, participants are also provided with pencils or drawing materials so they can illustrate their own ideas or complement their stories with visual representations, enriching the understanding of the use context from a more personal and creative perspective.

### 3.5 EMODIANA

EMODIANA is an instrument designed for the subjective measurement of emotions, based on the use of graphic representations that help users identify and express emotional states in an engaging and educational way. The tool includes a set of ten basic emotions: affection, joy, satisfaction, shame, sadness, boredom, seriousness, nervousness, surprise, and fear. Each of these emotions is represented by a distinctive character, visually designed to evoke the corresponding emotional expression [12].

In addition to emotional categorization, EMODIANA incorporates a color-coding system that allows users to indicate the intensity with which each emotion is experienced. By using specific colors, participants can express not only what emotion they are feeling, but also how strongly they feel it. This results in a richer, more nuanced understanding of their subjective emotional state. This instrument is especially useful in contexts where verbalizing emotions may be challenging, such as with children or individuals with communication barriers. It thus enables more inclusive and effective emotional assessment.

## 4 SESSIONS DESCRIPTION

As previously stated, the design of an accessible and motivating environment to engage the participation of users was planned. In this sense the motivation was around learning a new topic based on the Signwriting notation.

In this sense, a set of eight sessions distributed over two months were held at the school facilities. So, this gave us the opportunity to conduct research with Deaf users over multiple sessions.

To encourage consistent participation and avoid overburdening for both students and researchers, a schedule of two sessions per week was adopted. This approach ensured a steady workflow while allowing sufficient time for reflection, review of activities, analysis of preliminary results, and resolution of any challenges that might arise during the process.

The following sections provide a detailed description of each session, including the activities carried out in each one.

### 4.1 Session 1

During this first session, an introductory presentation on Signwriting was conducted with the group of students in the classroom. Colored gloves were used as a teaching resource to visually represent the colors used in the notation, thereby facilitating the association between movement and color, students paint their own hands to represent syntax, see **Figure 2**. In addition, illustrative drawings were created to demonstrate the basic principles of this writing system. Finally, a conceptual connection was established between the pictogram characteristic and the signs of Sign Language, allowing participants to understand the correspondence between these two forms of expression.

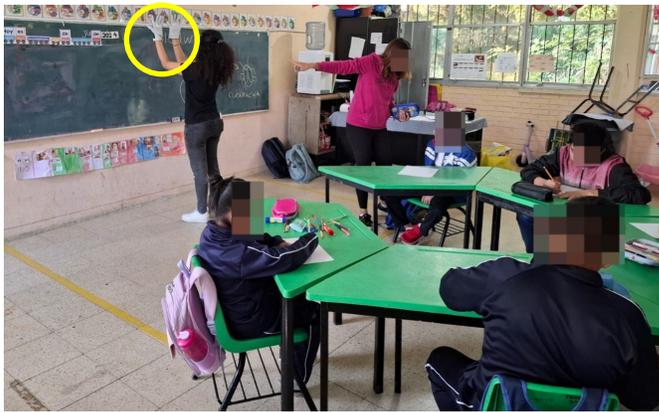 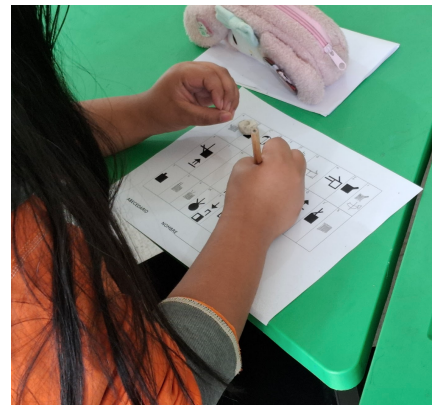

**Figure 2.** Globes (see yellow circle) and pictographic representations are used to introduce the notation. Source: Authors.

### 4.2 Session 2

In this session, practical activities were carried out to encourage exploration of the notation system. Black and white paint was applied to the children's hands, see **Figure 3**, which helped highlight the shape and orientation of the hands while producing signs. Afterwards, hand drawings were created as a strategy to deepen understanding of the graphic construction of signs. Specific signs in Sign Language—such as "cockroach," "hello," "doubt," "house," and "bicycle"—were practiced by the students and then transcribed into the notation. Finally, the colors used in the notation of these signs were identified and discussed, fostering a more detailed understanding of the system.

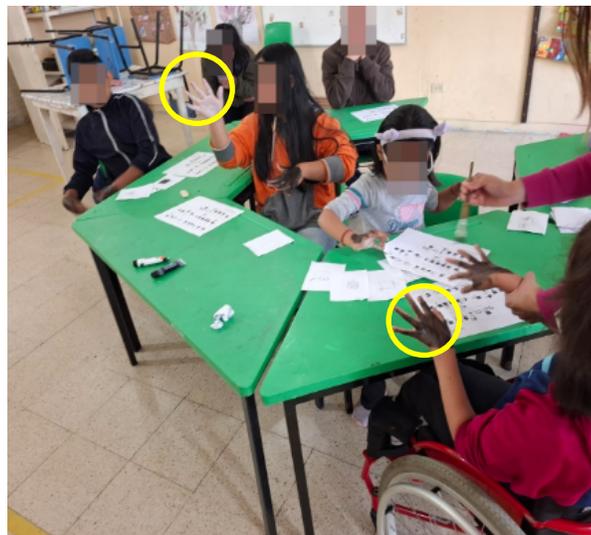

**Figure 3.** Students painting their hands (see yellow circles), and learn concepts about the notation. Source: Authors.

### 4.3 Session 3

An educational poster featuring the alphabet with the notation was presented. Using this material, students were invited to replicate the corresponding signs and place them on the whiteboard, see **Figure 4**. This activity allowed the children to actively engage with the symbols of the alphabet, reinforcing their learning through collaborative writing in the classroom.

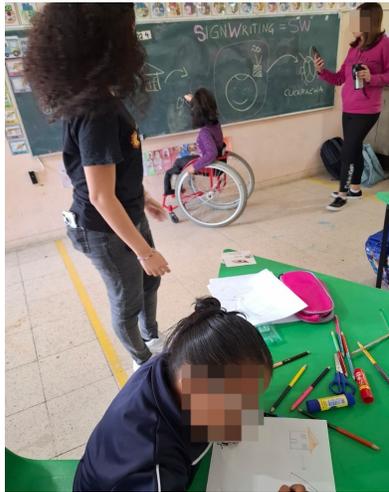

**Figure 4.** Active engagement collaborating on a whiteboard. Source: Authors.

## 4.4 Session 4

In this session, students were given access to the Signa App mobile application, see **Figure 5**. The children freely explored the various interfaces of the app, while the researchers conducted systematic observations of their interactions. Both gestures and participants' impressions and reactions were documented, with the aim of evaluating their user experience and identifying usability and accessibility aspects of the application.

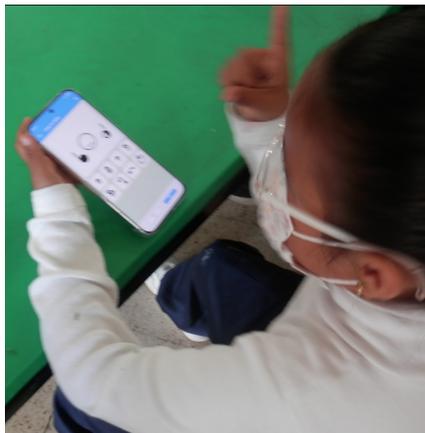

**Figure 5.** A student interacting with the app. Source: Authors.

## 4.5 Session 5

For this session, an activity was designed based on a memory matching game. Two sets of cards were prepared: one with images of animals and the other with their corresponding Signwriting notation. The children were tasked with matching each image to its written representation, thereby reinforcing the connection between the signs and their written form in a participatory and playful way.

### 4.6 Session 6

During this session, students once again used the Signa App mobile application. A specific task was assigned: to represent the sign "hello" using the notation within the app. The activity was carried out individually. The time each child took to complete the task was recorded, the device screen was captured to document their interactions, and notes were taken on any errors made. Afterwards, the users were asked to represent additional signs using the notation. This methodology allowed for an evaluation of both the app's design effectiveness and the end user's understanding of the system.

### 4.7 Session 7

In this activity, the children were given a sheet with the outline of a mobile device and cutouts of various interface elements from the application. They were also provided with pencils to draw freely. Using these materials, the students were invited to recreate the application from their own perspective, encouraging them to express their ideas and perceptions about the software's functionality and design.

### 4.8 Session 8

In this final session, free use of the Signa App mobile application was once again facilitated. At the end, the EMODIANA tool was applied, allowing the children to express the emotions they experienced while interacting with the app. A board featuring various illustrated emotional states (affection, joy, satisfaction, embarrassment, sadness, boredom, seriousness, nervousness, surprise, and fear) was used, enabling the collection of qualitative data about the users' emotional experience in relation to the application.

## 5 RESULTS

As previously discussed, five evaluation methods were selected to analyze the accessibility of each instrument applied. The results obtained were diverse, highlighting both strengths and weaknesses in the instruments. The process also demonstrated that not all traditional user experience evaluation methods are fully appropriate for Deaf individuals, as some do not adequately address their specific communicative and cognitive needs. These findings underscore the importance of adapting methodological approaches to the particular characteristics of users, especially in the context of accessibility.

### 5.1 User workshop

The implementation of a user workshop focusing on the Signwriting notation was proposed, because the notation was unknown to the students. Therefore the objective was to provide students with an adequate understanding of the fundamental concepts of it, enabling them to properly contextualize the use of the mobile application under study.

During the workshop, key principles of the notation were addressed, and various activities were carried out to promote its use as a tool for the written representation of sign language. The topics covered included:
- Sign language culture.
- The relationship between sign language and notation.

- The concept and foundational principles of the notation.
- The visual grammar and syntax specific to the notation.
- Analysis of examples and texts written in Signwriting through both group and individual activities.

By the end of the sessions, students were expected to understand how the notation works. In this context, it was expected to evaluate the usability of the application Signa App by verifying whether being oriented towards the use of sign language would make it easier to use.

Therefore, supported by the workshop technique, two groups of potential users—group A and group B—were engaged using different UX techniques. Additionally, the workshop proved to be an effective scenery by allowing integration with other complementary evaluation methods.

The results obtained from the workshop were positive, as evidenced by the activities carried out by the participants during the sessions. Children from both groups (A and B) successfully learned various concepts related to the notation, including the alphabet, animal names, days of the week, months of the year, and school supplies.

With the support of the teacher, the participants decorated the classroom with images and elements related to the notation, aiming to actively integrate them into their learning process. From another perspective, the workshop also allowed users to gradually become familiar with the functionality of the application.

## 5.2 Direct observation

This method was consistently integrated into all other evaluation techniques applied, as it served as a key tool for verifying and analyzing participants' reactions, skills, and feedback throughout the testing process. Its implementation allowed a detailed observation of how users interacted with the interface, helping to identify potential navigation difficulties, and enabled the collection of essential information regarding their user experience. In this regard, the method proved to be a fundamental component in ensuring the accuracy and depth of the results obtained at each stage of the evaluation.

To carry out this activity with the participants, five mobile devices were required, each one pre-installed with the Signa App application. Additionally, a camera was used to capture images that served as evidence of the activities conducted and the outcomes achieved. The researchers' planning included direct observation of user behavior, detailed note-taking on their reactions and actions, and the collection of visual material through photographs.

Before the evaluation through direct observation began, a complementary activity on the system notation was conducted. Its purpose was to help participants recall and reinforce the knowledge they had acquired during the user workshop. Then, with the assistance of an interpreter, participants were instructed to use the mobile application to generate pictograms.

Finally, they were encouraged to freely explore the application, navigating through all available interfaces and selecting all accessible options, in order to comprehensively assess their user experience and level of familiarity with the tool.

As a result, only four students actively participated in the navigation. One deaf participant decided not to complete the activity, primarily due to his temperament, which made him feel self-conscious when observed. This is because participation was assessed through direct observation, which made it possible to gather relevant information about user behavior during interface usage, as well as to identify potential areas for improvement of the app.

## 5.3 Picture card through prototyping

The primary purpose of this evaluation method was to develop a prototype draft that reflected the ideas and needs expressed by the users, with the aim of identifying and analyzing potential innovative features to be incorporated into a future version of the application. From a UX perspective, this exercise enables a schematic representation of the product interfaces and interaction flows, directly considering the users' perception and behavior during use.

To carry out this activity, each participant was provided with a blank sheet containing graphical representations of three mobile devices. Based on this template, users were asked to depict, through drawings, the main screens of the Signa App mobile application, focusing on the system's key functionalities. Throughout this process, participants had complete freedom to illustrate modifications or improvements they deemed appropriate, grounded in their experiences from previous usage sessions. Additionally, they were allowed to include extra cutouts, which could be affixed onto the paper prototype to further enrich their visual proposals.

Upon completion of the designs, each user shared their opinions and suggestions with the group, fostering a collaborative discussion space regarding the ideas presented. This exchange of perspectives facilitated the identification of relevant aspects for the continuous improvement of the product, as illustrated in **Figure 6.**

The ideas and explanations provided by users during the evaluation were diverse, reflecting a wide range of perceptions and expectations. Participants from group A primarily focused on aspects related to interaction during the process of creating pictograms. Some of them expressed that the interface felt monotonous and repetitive, which negatively impacted their user experience. In contrast, members of group B emphasized the need for personalization, particularly highlighting the possibility of creating customized avatars. From their perspective, these virtual representations would enable users to produce written signs in a more dynamic and meaningful way, fostering a more immersive experience. Although all users were able to complete the prototyping activity, several issues were identified during the evaluation process, including complications in the method's testing protocol and communication barriers, which will be analyzed in depth in the discussion section.

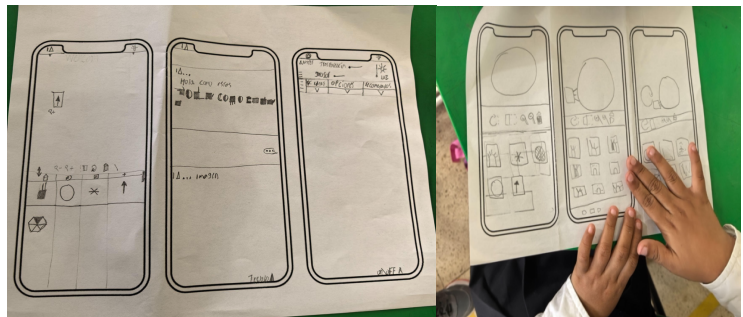

**Figure 6. D**esigns made by users. Source: Authors.

### 5.4  Cognitive walkthrough

This technique focused on analyzing how users perceive, understand, and make decisions during their interaction with the interface. This approach made it possible to identify the specific challenges faced by groups A and B while navigating through different sections of the application. It also enabled detailed observation of user-interface interactions, providing deep insight into the mental processes involved in navigating and using the system.

Prior to the cognitive walkthrough evaluation, a reinforcement activity about the system notation was conducted to recall the children the basic concepts of this. Subsequently, both groups were given access to the Signa App mobile application and were assigned specific tasks designed to assess their interaction with the interface.

The tasks involved observing a predefined model and accurately replicating it using the tools available in the interface, in **Figure 7** is shown the replica task. The specific tasks assigned to the participants are detailed below.

- **Task 1:** generate the pictogram for "House", the estimated time is 5 minutes.
- **Task 2:** generate the pictogram for "Bell", the estimated time is 5 minutes.
- **Task 3:** generate the pictogram for "Cat", the estimated time is 5 minutes.
- **Task 4:** generate the pictogram for "Name Sign", which corresponds to the participant's identifying nickname (in sign language), the estimated time is 8 minutes.

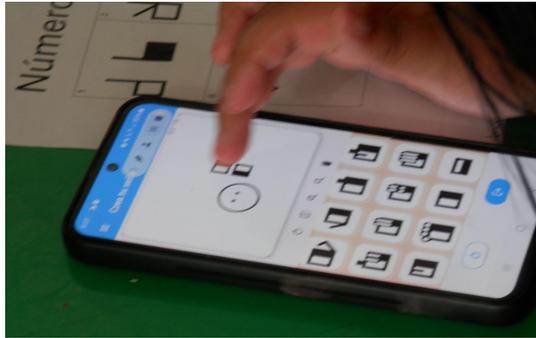

**Figure 7.** Replica of the Pictogram "Bell". Source: Authors.

Overall, all users in groups A and B were able to complete the assigned activities properly and without the need for external assistance. In **Table 5** it can be seen the time performed to complete the tasks. Based on the results obtained, it was decided to investigate whether there were significant differences in performance between the two groups. Thus, two statistical analyses were performed: Welch's t-test for quantitative data and Fisher exact test for qualitative data. Welch's t-test was conducted according to equations (1) and (2). This method accounts for differences in variance and sample size between groups. The results of the analysis are presented in **Table 6**.

**Table 5.** Cognitive walkthrough results expressed in seconds for completion tasks. Source: Authors.

| Group | User | Task 1 | Task 2 | Task 3 | Task 4 | Average |
|-------|------|--------|--------|--------|--------|---------|
| A | 1 | 244 | 271 | 248 | 437 | 300 |
| A | 2 | 268 | 338 | 265 | 198 | 267.25 |
| A | 3 | 383 | 427 | 403 | 408 | 405.25 |
| B | 4 | 361 | 327 | 283 | 208 | 294.75 |
| B | 5 | 387 | 324 | 289 | 402 | 350.5 |

$$t = \frac{\bar{x}_1 - \bar{x}_2}{\sqrt{\frac{s_1^2}{n_1} + \frac{s_2^2}{n_2}}} \quad (1)$$

$$df = \frac{\left(\frac{s_1^2}{n_1} + \frac{s_2^2}{n_2}\right)^2}{\frac{\left(\frac{s_1^2}{n_1}\right)^2}{n_1 - 1} + \frac{\left(\frac{s_2^2}{n_2}\right)^2}{n_2 - 1}} \quad (2)$$

**Table 6.** Independent samples t-test assuming unequal variances. Source: Authors.

|  | Group A | Group B |
|---|---|---|
| Mean | 324.17 | 322.63 |
| Variance | 5199.02 | 1554.03 |
| Number of observations | 3 | 2 |
| Hypothesized mean difference | 0 | - |
| Degrees of freedom (df) | 3 | - |
| t statistic | 0.0308 | - |
| p-value (one-tailed) | 0.4887 | - |
| Critical t-value (one-tailed) | 2.3534 | - |
| p-value (two-tailed) | 0.9774 | - |
| Critical t-value (two-tailed) | 3.1824 | - |

Given that the **p-value = 0.9774** is much greater than the standard significance level of 0.05, we **fail to reject the null hypothesis**. This suggests that **there is no statistically significant difference** in the mean task completion times between group A and group B.

The success completion tasks for each user and groups is shown in **Table 7**. With value 1 to indicate success and value 0 to indicate no success.

**Table 7.** Cognitive walkthrough results expressed by completion tasks. Source: Authors.

| Group | User | Task 1 | Task 2 | Task 3 | Task 4 | Completed |
|---|---|---|---|---|---|---|
| A | 1 | 1 | 1 | 1 | 1 | 4 |
| A | 2 | 1 | 0 | 1 | 1 | 3 |
| A | 3 | 0 | 0 | 0 | 0 | 0 |
| B | 4 | 0 | 0 | 1 | 1 | 2 |
| B | 5 | 0 | 0 | 1 | 1 | 2 |

To determine whether there was a significant association between user groups (A or B) in task completion, a Fisher's Exact Test was conducted on a 2×2 contingency, shown in **Table**

8. This test is appropriate for small sample sizes and evaluates whether the observed distribution differs from what would be expected under the null hypothesis of independence. The formula for this operation is shown in equation (3).

$$P = \frac{\binom{a+b}{a}\binom{c+d}{c}}{\binom{n}{a+c}} \tag{3}$$

Table 8. Contingency table for task completion in both groups. Source: Authors.

|  | Completed | No Completed |
|---|---|---|
| Group A | 7 | 5 |
| Group B | 4 | 4 |

Since the p-value equals 1.0, we fail to reject the null hypothesis. There is **no statistically significant difference** in task completion rates between Group A and Group B. The difference in proportions (58% vs. 50%) is not sufficient to infer an effect with the current sample size.

Finally, some general observations regarding the application include:
- Suggestions were made to organize the pictograms based on their colors rather than their configuration.
- It was decided to create "Dots" to keep the user informed about their current level within the application.
- The arrangement of pictograms that were placed at incorrect levels was corrected.
- A higher level of interface adoption was observed among members of group A compared to those in group B.

### 5.5 EMODIANA

The purpose of this evaluation was to identify the emotions experienced by users during their interaction with the application. To this end, the EMODIANA instrument was employed, which enables participants to express their emotions using an intensity scale, thereby facilitating the understanding of their emotional state throughout the app usage. Prior to implementing this activity, a planning session was conducted in collaboration with the teacher, as it was necessary for the children to become familiar with the various emotional states represented in the instrument. For this purpose, sensitization activities were carried out using illustrative posters in the classroom, which addressed a range of emotions.

After receiving a reminder about the different types of emotional states, the children were given mobile devices so they could interact autonomously with the application. Subsequently, they were presented with the EMODIANA board, which displayed ten distinct emotions, each represented by an avatar. The intensity of each emotion was indicated using a color-coded scale.

However, the evaluation could not be successfully completed, as the participants were unable to finalize their use of the EMODIANA board. Although this instrument was specifically designed to be understood by children, significant difficulties were observed in both evaluated groups (group A and group B). The accessibility of the instrument emerged as a critical issue and will therefore be subject to detailed analysis. Several factors were identified that may have contributed to the challenges experienced by users during the application of the EMODIANA method (see them in discussion section), which will be further analyzed in the Discussion section.

For this research, as well as for the joint development of the application used, the method was adapted to reduce emotions to only seven categories: joy, sadness, boredom, shame, surprise, fear and anger, see **Figure 8**. It is noteworthy that users interpreted joy and boredom as "Liking" and "Disliking," respectively. Furthermore, it was taken into account that, when working with deaf people, it is essential to use clear, simple, and direct language.

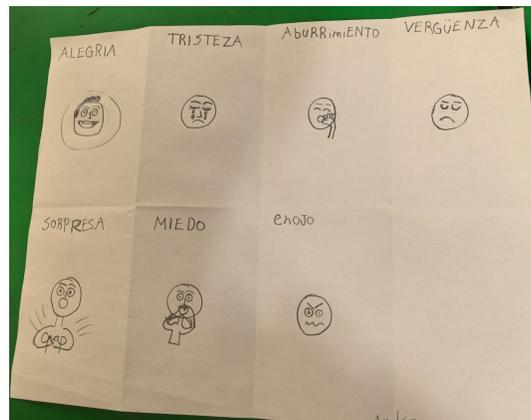

**Figure 8.** Emotions recognized by Users. Source: Authors.

# 6  DISCUSSION

This section analyzes the accessibility of the various UX evaluation methods employed in this study when applied to Deaf individuals. The main challenges and limitations faced by these participants during the evaluation processes are identified and examined, with particular attention to factors that may hinder an accurate and effective assessment of their experience. The objective is to highlight how certain communicative, methodological, or contextual barriers may negatively impact the validity of the results, and to reflect on the need to adapt or redesign evaluation methods to ensure inclusion and equity in user experience research.

## 6.1  Limitations

Although the evaluation methods discussed have been previously recommended in the literature for use with Deaf individuals, their concrete implementation with this population revealed several limitations. These challenges became apparent during the evaluation sessions, where it was observed that the linguistic and cultural specificities of Deaf participants had not been adequately considered. This oversight negatively affected the effectiveness of the applied methods. The following section will examine in detail the limitations observed during the pilot sessions conducted with groups A and B, with the aim of critically analyzing the problematic aspects and identifying opportunities for improvement.

### 6.1.1  User workshop

While this evaluation method offers the advantage of allowing the simultaneous participation of a considerable number of users, as well as the parallel execution with other evaluation methods, several limitations were identified during its implementation.

As a first consideration, conducting a user workshop involves organizing multiple sessions specifically tailored to the participants. This structure addresses the need for users to acquire the fundamental prior knowledge that enables adequate familiarization and contextualization with the content and functionalities of the application under evaluation.

However, irregular attendance was observed among users in the scheduled activities, despite the implementation of various strategies and resources aimed at promoting active participation in the study. This situation necessitated repeating several sessions and activities related to learning the proposed notation system, which not only considerably extended the planned session durations, but also caused some discomfort among certain participants. For some, the repetition of content was not perceived as a pedagogical reinforcement but rather as an unnecessary interruption that led to frustration and dissatisfaction within the group—an effect attributable to the nature of the child population involved in the study.

As a result, the original activity schedule was disrupted, and the evaluation calendar had to be substantially modified. Various unforeseen circumstances forced the rescheduling of several tests to ensure that all users had the opportunity to fully participate in each evaluation instance. Despite the irregular attendance, all activities planned for the user workshop were completed, thereby meeting each of the established objectives.

### 6.1.2 Direct observation

Participants from both groups proceeded to use the application, allowing observation of their interaction with the proposed interfaces. In the case of Deaf users, it was evident that each participant employed a variety of gestures, signs, and facial expressions to communicate their emotions and ideas. To properly interpret these manifestations, the intervention of the teacher was essential, as they facilitated mediation and comprehension of the feedback provided by the users. In this regard, if the observer lacks sufficient knowledge of sign language or is unfamiliar with the cultural and communicative variations within the Deaf community, there is a considerable risk of misinterpreting the observed behaviors, which could compromise the accuracy and validity of the evaluation.

In addition, one of the Deaf participants expressed confusion during the assessment due to uncertainty about how to use the application. This situation required direct intervention from the researchers, who provided guidance and encouragement to help the participant interact effectively with the tool. However, this assistance led the participant to experience a sense of mistrust, which may have influenced their responses and introduced biases that affect the reliability of the results obtained.

Finally, it is essential to recognize that the Deaf community has its own culture, characterized by particular forms of interaction, values, and specific norms. Therefore, direct observation without appropriate cultural context may lead to misinterpretations and misunderstandings. This underscores the importance of involving trained evaluators who are culturally sensitive and knowledgeable about these particularities, in order to ensure accurate and respectful evaluations. The key findings from the user evaluation are presented below.

### 6.1.3 Cognitive walkthrough

The task instructions through Sign Language proved to be insufficient for Deaf users to fully comprehend the activities. This led to confusion and delays in carrying out the tasks, as participants struggled to effectively grasp all the key aspects of the assigned activities. To address this challenge, an alternative strategy was adopted to enhance comprehension. A whiteboard was used to illustrate the various steps required to complete the activities through drawings. This visual method allowed users to clearly see each required action, thereby facilitating a better understanding of the instructions. The use of graphic representations not only clarified the purpose of each task but also significantly contributed to more efficient and accurate execution by the participants.

During one of the tests, a Deaf user expressed confusion while navigating the interfaces, which led them to stop and cease interaction with the application. In response, the evaluators intervened to provide assistance and address the questions that arose during the session. They took the necessary time to explain in detail the various functionalities offered by the application, while also ensuring that the user felt comfortable and understood each of the required steps. However, this intervention influenced the evaluation results, as it prevented the user's interaction with the application from being entirely natural and spontaneous, potentially introducing bias into the collected data.

### 6.1.4 Picture card through prototyping

In the course of the evaluation, multiple challenges affecting effective user participation were identified, particularly within group A. One Deaf participant, with limited knowledge of MSL, struggled to understand the proposed activity. As a result, rather than generating new design alternatives for the application, they replicated the existing interfaces exactly. This situation highlights the barriers inherent in the evaluation methods used, which may compromise both accessibility and the accuracy and reliability of the data obtained.

Additionally, feedback was collected regarding the application, including a Deaf user who expressed that the interface seemed boring and suggested some improvements. However, due to limitations in their proficiency with certain specific signs, they were unable to clearly convey their ideas, which complicated the researchers' interpretation of their suggestions. This type of limitation underscores the urgent need to adapt evaluation methods to ensure an inclusive experience that more accurately represents the perceptions, needs, and proposals of Deaf users.

### 6.1.5 EMODIANA

This method is especially distinguished for its applicability to child populations, as it offers an innovative approach to the subjective measurement of emotions and their intensity in children. The tool has been widely recognized for its effectiveness in capturing emotional responses, providing relevant insights into users' affective states during their interaction with a product [12]. However, when applied to both groups of users, with and without hearing impairments, significant difficulties in completing the evaluation properly were observed. These findings suggest that, despite being designed for a child audience, the instrument presents limitations in terms of accessibility and comprehension—especially when considering the diversity of communicative abilities within the evaluated group.

Once both groups had completed their interaction with the application, the users showed a marked interest in the avatars presented on the EMODIANA board. However, these avatars acted as a source of distraction, as the children began imitating the emotional expressions portrayed by the characters, diverting their attention from the main task. Additionally, some participants took sheets of paper to manually draw each of the avatars, labeling them with the corresponding emotion names. This behavior indicates a high level of fascination with the avatars, which may have contributed to increased distraction and hindered the proper application of the evaluative method.

Regarding the colors on the board—designed to indicate the intensity of each emotion—these went largely unnoticed during the evaluation process. The concept of intensity proved to be complex for the users, despite the explanations provided by both the teacher and the evaluators. This difficulty highlights that the association between colors and the level of emotional intensity was neither sufficiently clear nor accessible for this child population, which negatively affected the instrument's ability to effectively measure emotions and their intensity.

## 7 CONCLUSION

This research addressed the analysis of user experience evaluation methods from an accessibility perspective, focusing specifically on their application during software testing with deaf individuals. The primary objective was to assess the suitability of the instruments used to collect meaningful data during UX evaluations with this population, as well as to identify the limitations arising from their implementation—particularly those related to the communicative, cognitive, and cultural characteristics of the deaf community.

The results demonstrated that, while it is possible to adapt traditional UX evaluation methods to include deaf participants, their effective application requires substantial modifications. These adaptations must explicitly consider the linguistic, communicative, and cultural specificities of this community in order to ensure not only their participation, but also the validity of the data collected. Throughout the testing sessions, significant limitations were identified in each of the methods used, affecting both the quality of the information obtained and the overall experience of the participants.

Specifically, the user workshop, although useful for providing context to participants, presented logistical and pedagogical challenges related to irregular attendance and the negative perception of repetitive content. Direct observation revealed potential biases and misinterpretations when observers were unfamiliar with Mexican Sign Language and deaf culture. The cognitive walkthrough showed that instructions delivered in MSL did not guarantee full comprehension, necessitating additional visual supports such as whiteboards.

In the illustrated card method used during prototyping, linguistic barriers limited participants' ability to express their ideas, which affected the quality of the proposals gathered. Finally, the use of the EMODIANA instrument revealed that, despite the visual appeal of the emotional avatars, they became a source of distraction from the main task, and the concept of emotional intensity was not fully understood, complicating the interpretation of the results.

Based on the findings, it is concluded that specific adaptations to the evaluation methods are essential to ensure true accessibility for deaf individuals. These modifications will not only support more equitable participation but also lead to more accurate reflections of this population's experiences, perceptions, and needs.

Finally, the results of this study reveal the need to design and adapt UX evaluation methods that are not only linguistically accessible, but also aligned with and respectful of the cultural identity of the deaf community. Achieving this requires the involvement of accessibility specialists from the earliest stages of methodological design and the participation of Deaf community. Following this approach more inclusive and representative evaluations can be developed. This work not only highlights the current challenges in this area but also lays a solid foundation for the development of more inclusive evaluative methods within the field of user experience. Furthermore, it emphasizes the importance of continuing to explore new mechanisms and approaches that broaden the scope of UX evaluation, thereby contributing to the creation of more accessible and equitable digital environments for all users.

In future work, it is considered necessary to expand the sample size and extend the study to different age ranges, where other types of barriers and specific needs will surely be identified.

## DECLARATION OF COMPETING INTEREST

The authors declare that they have no known competing financial interests or personal relationships that could have appeared to influence the work reported in this paper.

## Credit authorship contribution statement

Andrés Eduardo Fuentes-Cortázar: Writing – original draft, Writing – review & editing, Data curation.
Alejandra Rivera-Hernández: Writing – review & editing, Software, Data curation.
José Rafael Rojano-Cáceres: Writing – review & editing, Formal analysis, Conceptualization, Supervision.